\documentclass[12pt,preprint]{aastex}
\usepackage{graphicx}

\shorttitle{Partial eruption of a prominence}
\shortauthors{Durgesh Tripathi et al.}

\begin{document}

\title{SDO/AIA Observations of a Partially Erupting Prominence}
\author{Durgesh Tripathi}
\affil{Inter-University Centre for Astronomy and Astrophysics, Pune University Campus,  Pune 411007, India}
\author{Katharine K. Reeves}
\affil{Smithsonian Astrophysical Observatory,60 Garden street, MS 58, Cambridge, MA 02138, USA}
\author{Sarah E. Gibson}
\affil{High Altitude Observatory, National Center for Atmospheric Research, Boulder, USA}
\author{Abhishek Srivastava and Navin C. Joshi}
\affil{Aryabhatta Research Institute of Observational Sciences (ARIES), Nainital 263129, India}

\begin{abstract}We report an observation of a partially erupting prominence and associated dynamical plasma processes based on observations
recorded by the Atmospheric Imaging Assembly (AIA) on board the Solar Dynamics 
Observatory (SDO). The prominence first goes through a slow rise (SR) phase followed by a fast 
rise (FR). The slow rise phase started after a couple of small brightenings seen toward the 
footpoints. At the turning point from SR to FR, the prominence had already become kinked. The prominence 
shows strong brightening at the central kink location during the start of FR. We interpret this as internal magnetic 
reconnection occurring at a vertical current sheet forming between the two legs of the 
erupting prominence (flux-rope). The brightening at the central kink location is seen in all the 
EUV channels of AIA. The contributions of differential emission at
higher temperatures are larger compared to that for typical coronal 
temperatures supporting a reconnection scenario at the central kink location.
The plasma above the brightening location gets ejected as a hot plasmoid-like structure embedded in a CME, and 
those below drain down in the form of blobs moving towards the Sun's surface. The unique time 
resolution of the AIA has allowed all of these eruptive aspects, including SR-to-FR, kinking, 
central current sheet formation, plasmoid-like eruption, and filament ``splitting'', to be observed in a single event, 
providing strong and comprehensive evidence in favour of the model of partially erupting flux 
ropes. \end{abstract}


\section{Introduction}

Prominence eruptions are one of the best proxies for Coronal Mass Ejections (CMEs). 
Therefore, understanding the initiation and evolution of the erupting prominences provides 
crucial physical understanding of the dynamical processes involved in CME initiation and 
evolution, with broader implications for space weather and geo-space climate.

Prominence eruptions show a range of observational characteristics from complete eruption to 
partial eruption to failed or sympathetic eruption. Partial eruption of prominences is a relatively 
new phenomena which was first described by \cite{gilbert_2000}. These partial eruptions were 
observed in form of downflows from the erupting prominences. \cite{gilbert_2000, gilbert_2001} 
found that majority of the erupting prominences observed with H$\alpha$ coronagraph located 
at the Mauna Loa Solar Observatory (MLSO) were associated with downflows. Based on these 
observations \cite{gilbert_2001} hypothesised that during the process of eruption the 
prominence breaks into two parts involving an X-type neutral point and internal magnetic 
reconnection. Depending on the location of internal magnetic reconnection the prominence 
would show complete, partial or failed eruption \citep{gilbert_2001}.

\cite{gibson_2006_apj, gibson_2006_jgr} studied evolution of a flux-rope from the Sun's surface 
out to 6 solar radii where a flux-rope with enough twist erupts due to loss of equilibrium.  In this 
simulation a flux-rope undergoes a kink instability leading to the formation of a vertical current 
sheet between the two legs of the flux-rope \citep[see also ][]{birn_2006}. Based on these 
simulations, \cite{gibson_2006_apj, gibson_2006_jgr} demonstrated that multiple internal 
reconnections (inside the flux-rope) take place at the vertical current sheet. This leads to 
the bifurcation of the flux-rope. In  this process, the part of the flux-rope above the reconnection 
point escapes as the core of CME and the part of the flux-rope below the reconnection point 
falls back towards the Sun's surface. \cite{SriZ:10} have found observational evidence of kink instability 
in a highly twisted loop system and the formation of a horizontal current sheet below the double fragmented loop-top.

Based on the above mentioned simulations, \cite{gibson_2006_jgr} predicted 
some observables which could be directly compared with observations. \cite{tripathi_2009} 
studied 6 CMEs events associated with erupting prominences/filaments and concluded that 
the model of \cite{gibson_2006_apj, gibson_2006_jgr} involving partial eruption of flux-rope during the CME 
eruption was capable of producing almost all the observed signatures in different CME events. Thus, the general 
behaviuor described by the model is broadly applicable to partial eruptions, assisting in interpretation of the CME 
evolution and its interplanetary consequences.

\cite{alex_2006} reported a prominence eruption using the Transition Region and Coronal 
Explorer \citep[TRACE; ][]{trace} observations together with Reuven Ramaty High Energy Solar 
Spectroscopic Imager \citep[RHESSI;][]{rhessi} observations. They identified two coronal hard 
X-ray sources, which were located above the filament before the main activation phase 
\citep[see also][]{ji_2003} and right under the apex of the filament when the filament was strongly 
kinked. \cite{alex_2006} attributed this second brightening to an internal magnetic reconnection 
occurring at the vertical current sheet formed under the apex of the kinked filament slightly above the 
projected crossing point of its two legs. This 
event was classified as a failed eruption as there was no associated CME observed.
\cite{liu_2007} reported an observation of partial eruption where the filament/cavity 
structure bifurcated into two parts with the upper part ejected as a CME and the lower part 
along with the bulk of the filament material remaining on the Sun's surface. This event was 
classified as a case of partially erupting prominence. However, no clear evidence of 
reconnection was observed related to this event. 

Using multi-wavelength observations recorded by Extreme-ultraviolet Imaging Telescope (EIT) 
aboard Solar and Heliospheric Observatory (SOHO), Soft X-ray Telescope (SXT) aboard 
Yohkoh, and H$\alpha$ and white-light coronagraphic observations from the MLSO, 
\cite{tripathi_2006, tripathi_2007} reported observations of a coronal downflow in the wake of a 
prominence eruption which was bright in EIT and XRT observations. The speed of the downflow 
measured using the while-light coronagraphic data was equivalent to the local Alfv\'en speed. 
Furthermore, a cusped shaped structure was observed in SXT images. The downflow started at 
the location of X-ray cusp and a base difference images showed an increase in the area of the 
dimming region at the cusp. All these observed features were interpreted as signatures of magnetic 
reconnection occurring at the location of the cusp. However, due to different limitations to the observations, 
the complete evolution of 
the prominence eruption was not observed. For instance, the time resolution of EIT or SXT was not enough 
to show if the prominence exhibited kinking motions. In addition, neither EIT nor SXT observed the 
apex of the cusp because of the limited field of view. On the other hand, H$\alpha$ and white-light  observations 
at MLSO showed the apex and location of the bifurcation of the erupting 
prominence, but were not available for the early phase evolution, and, since white-light observations
are temperature independent they could not provide information about the thermal properties of the plasma.

The six events studied by \cite{tripathi_2009} showed many characteristics predicted by the 
model of \cite{gibson_2006_jgr}. These events separately showed different predicted observables, but none of the 
events had all the predicted observables together. In addition, due to limited temporal and spatial resolution, 
some of the predicted characteristics could not be studied, in particular a transition from slow rise (SR) to fast rise (FR), 
kinking, central current sheet formation, and filament ``splitting''. 

In the present paper we report an observation of a partially erupting prominence and 
associated dynamics of the hot plasmoid-like structure using the observations recorded by 
the Atmospheric Imaging Assembly (AIA) onboard the Solar Dynamics Observatory (SDO). The spatial and temporal 
resolution of AIA observations and broad temperature coverage is ideal for such observations. In particular, we are 
able  to study the complete evolution and dynamics of an erupting prominence and detect eruptive aspects indicative 
of a partially-ejected flux rope in the observations. The remainder of the paper is organised as follows. In section~\ref{obs} 
we describe the observations followed by the data analysis and results in section~\ref{analysis}.  We summarise our 
results and discuss in section~\ref{conc}.

\section{Instrumentation and Observations} \label{obs}
A small erupting prominence was observed on 12 Sep 2011 by AIA at the East limb of the Sun in a region later named as 
NOAA~AR~11295 on 13 September. AIA provides observations of the Sun in eight different wavelength channels
simultaneously with a typical cadence of 12 seconds and pixel size of  0.6 arc sec. The primary 
contributing ion and the temperature response for each of these channels can vary depending on the 
features on the Sun's surface being observed. For more details see \cite{brendan, giulio_sdo}. In this paper we 
have investigated all the EUV channels with our main emphasis on 304~{\AA}, 193~{\AA} and 211~{\AA}, and
have focussed on the early-phase evolution of the prominence and plasma heating.

Since this is an event occurring on the limb of the Sun, we have also used the observations recorded using SECCHI 
\citep[Sun Earth Connection and Heliospheric Investigation;][]{secchi} onboard STEREO-B (Solar Terrestrial Relation 
Observatory), which, because of specific orbit, provides complementary on disk information for the same event. At the time of 
the observations presented here, the location of STEREO-B spacecraft was such that the features which were seen by SDO 
at east solar limb, STEREO would see them nearly at the disk centre. Therefore, it provides complementary vantage point for observations.

\section{Data Analysis and Results}\label{analysis}

\begin{figure}
\centering
\includegraphics[width=1.0\textwidth]{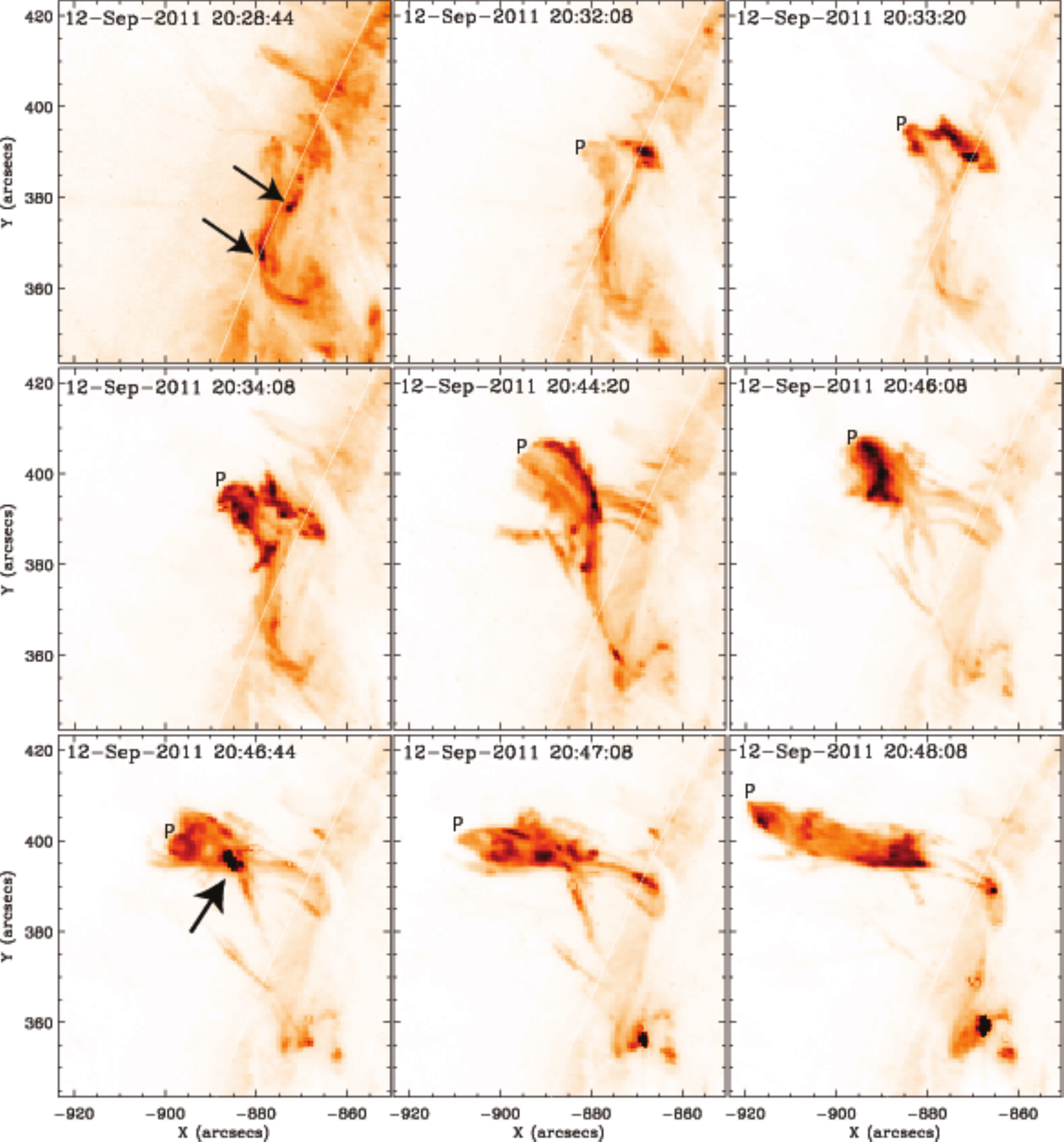}
\caption{Sequence of AIA images showing the erupting prominence taken in 304~{\AA} passband. Two arrows in the top left image locate the first two 
brightenings. 'P' locates the feature which was tracked to obtain the height-time diagram shown in Fig.~\ref{ht}. The arrow in the bottom left panel 
located the intense brightening at the location of the kink. \label{aia304}}
\end{figure}

Figure~\ref{aia304} and the accompanying animation shows a sequence of images taken in 304~{\AA},  passbands of AIA. 
Note that the images are displayed in negative intensity. The prominence stays relatively quiet until 20:28:44 UT, when two distinct 
brightenings are seen, marked by two arrows in the top left panel, one near the upper leg of the prominence and the other near the 
middle. We scrutinised the observations taken in the other bands of AIA and found a similar morphology. The two brightenings 
(marked by arrows) are strongest in 304~{\AA} channel but also seen mildly in 193~{\AA} and to a lesser extent in 211~{\AA}.

The complete prominence brightens up at around 20:30 UT, about 2 minutes after the first two brightenings, 
as seen in the 304 images. These images clearly show that the prominence is comprised of many different 
threads (strands). The brightening in the northern footpoint increases and the prominence goes thorough 
morphological evolution as revealed by the high time resolution of AIA images. During this time the 
prominence shows a SR phase. Similar features have previously been reported by \citet{chifor:2006, chifor:2007, it_2006}. 
At 20:40:08 UT (after about 12 mins), when the prominence has 
attained some height, a large part of the prominence brightens up. At 20:43:20~UT the morphology of the 
prominence resembles of a kinked flux rope (slinky) with two footpoints attached to the surface. A similar kind of 
morphological behaviour is seen in the other EUV passbands of AIA. The kinked prominence continues to rise. 
The prominence seems to go through heating and cooling as manifested by many localised brightenings that again
indicate thermally independent strands comprising the prominence. By 20:46:08 UT (after about 3 minutes of the kink), most of 
the plasma located above the kink location becomes much brighter than the two legs of the prominence that 
are still connected to the surface (see middle right panel in Fig.~\ref{aia304}). 

\begin{figure}
\centering
\includegraphics[width=1.0\textwidth]{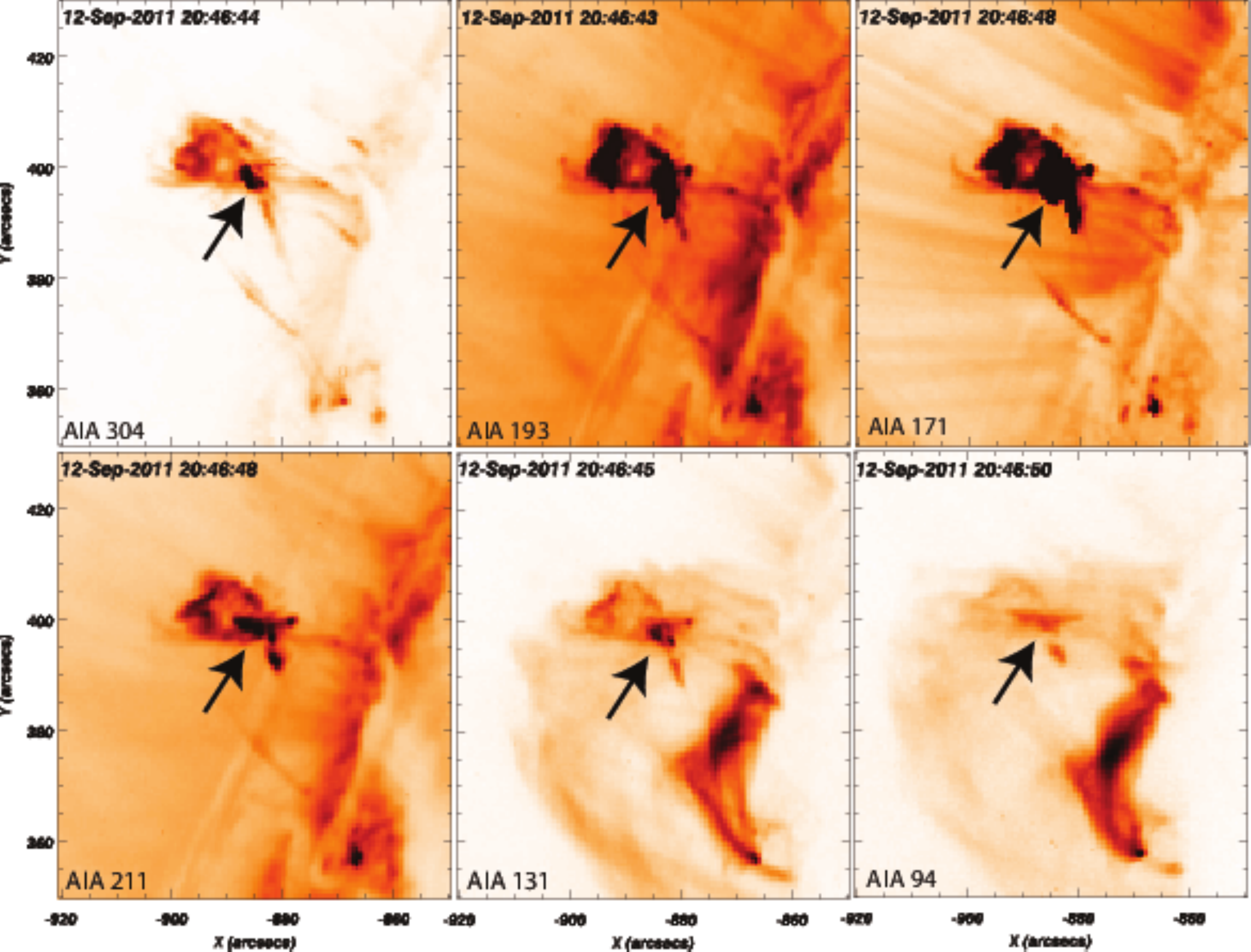}
\caption{Sequence of AIA images recorded with its all the EUV channels showing the brightening at the location of kinked flux-rope.\label{aia_all}}
\end{figure}

At around 20:46:32 UT (see animation) and 20:46:44 UT (shown by an arrow in bottom left panel in Fig.~\ref{aia304}), brightenings 
took place exactly at the location of the kink. Figure~\ref{aia_all} shows near simultaneous observations taken using all six EUV 
bandpasses of AIA. The brightening shown in the bottom left panel in Fig~\ref{aia304} is located by an arrow in all the EUV channels of 
AIA in Fig.~\ref{aia_all}. This sudden brightening at the kinked location, which is seen in cooler as well 
hotter channels of AIA, is strongly suggestive of magnetic reconnection occurring at the kink location at the current sheet which may have 
formed between the two legs of the erupting flux-rope. 

\begin{figure}
\centering
\includegraphics[width=0.65\textwidth]{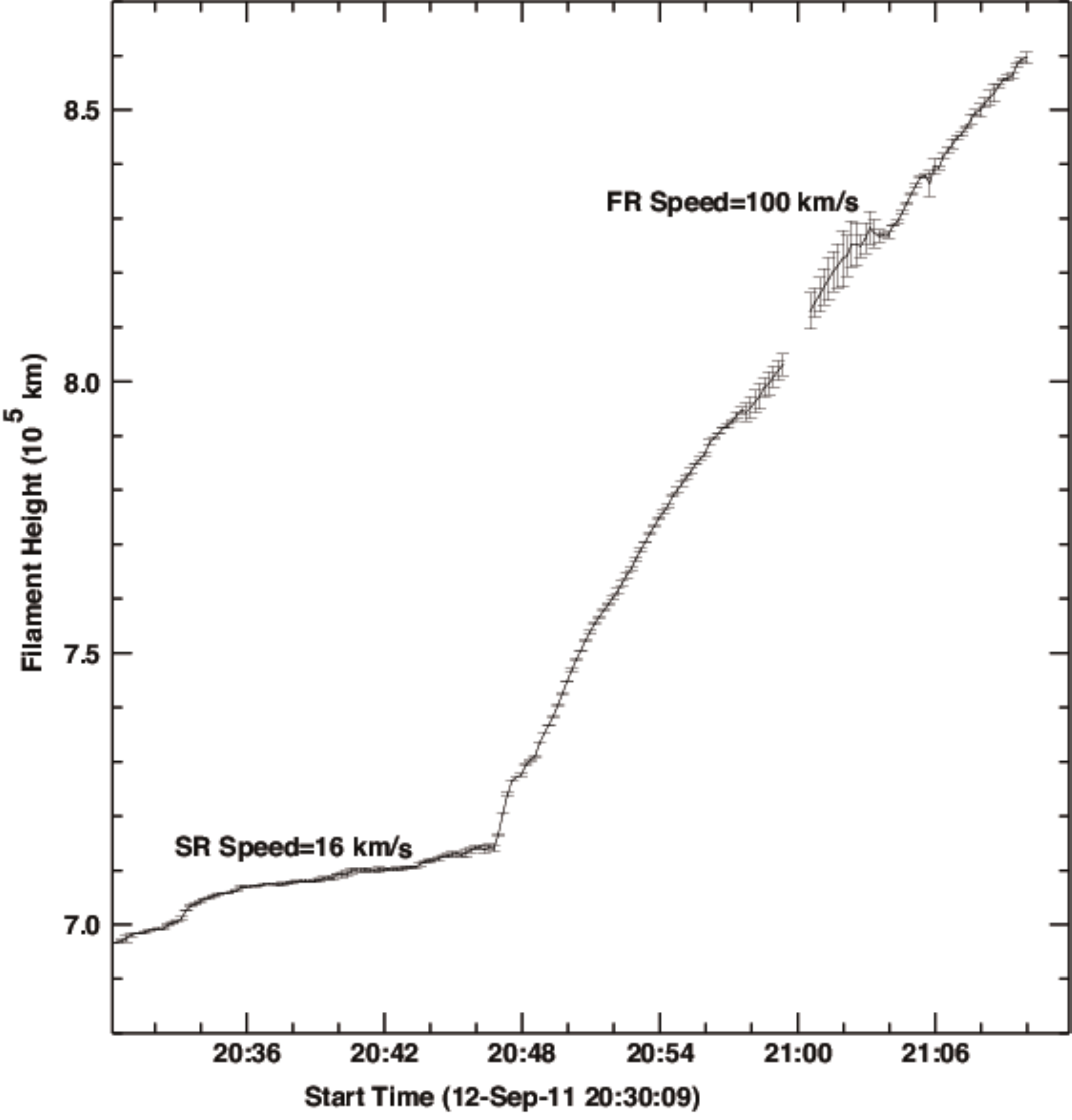}
\caption{Height-time plot of the erupting prominence measured using the AIA 304~{\AA} observations. This height-time plot was generated by tracking the feature denoted by 'P' in Figure~\ref{aia304}.\label{ht}}
\end{figure}

Figure~\ref{ht} displays the height-time plot for the erupting prominence measured using the observations recorded in the 304~{\AA} 
channel. The plot was obtained by tracking the tip of the erupting prominence. The error bars are due to the standard deviation of 
three simultaneous measurements of the same features. The plot clearly demonstrates that the prominence goes through a SR
phase up till  $\sim$20:47~UT before it is accelerated. A linear fit to the data point from the beginning till 20:47~UT yields a speed of 
about 16~km~s$^{-1}$ and a separate linear fit to the data from 20:47~UT onwards yields a speed of $\sim$98~km~s$^{-1}$. The 
height time plot show a clear 'knee' at around 20:47~UT, which matches with the time of the kinking of the prominence.

We note that the SR to FR change occurs around 20:47 UT, while the kinking becomes
evident somewhat earlier around 20:44 UT. Instead of the sharp turning point, there is a knee (cf., Fig.~3) observed 
that indicates that the kinked prominence threads involved in a bulk reconnection during 
a span of the time between 20:44-20:47 UT and thereafter prominence rises abruptly. The plasma which is above the location of the 
kink are ejected outwards and that below drains towards the 
surface, similar to the observations reported earlier \citep[][see also \cite{gilbert_2001}]{tripathi_2006, tripathi_2007}.
The associated CME seen in LASCO/C2 FOV (not shown) 
first appear at 20:12 UT and disappear quickly within few frames of LASCO/C2 observations. This supports the scenario of reconnection at a 
central kink location and splitting of the 
prominence in a partial eruption. Unfortunately, we do not have hard X-ray observations by RHESSI
in the similar time domain of the event, therefore, we cannot examine the formation of the HXR
at the central kink location as previously observed in various studies \citep[][]{Tetal_2004, alex_2006, liu_2009}.

In addition, there are signatures of interaction between the eruptive filament system and the overlying  loops, as reflected 
in form of de-acceleration in the height time diagram shown in Fig. 3. We also note that the overlying fields play important role in 
prominence eruptions as shown by e.g., \cite{tk_2005, fg_2007, kli_2012}.

\begin{figure}
\centering
\includegraphics[width=0.85\textwidth]{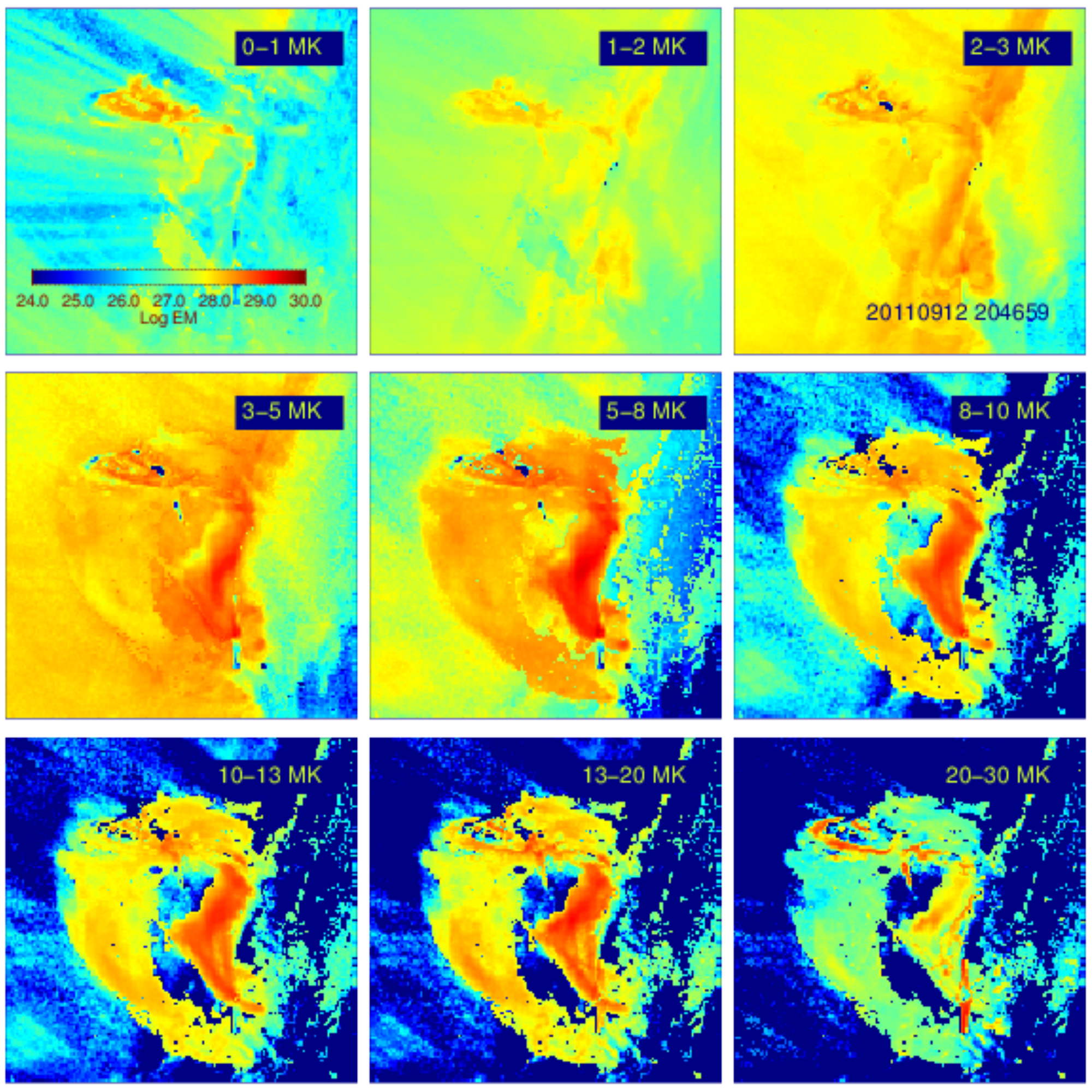}
\caption{Emission measure maps obtained using six channels of AIA at 20:46:59 UT.\label{em}}
\end{figure}

In order to understand the thermal structure of the erupting prominence and possibly the associated flux rope we have derived 
emission measure (EM) maps in different temperature bins (see Figure~\ref{em}) using the six AIA channels. DEMs were calculated 
using the \textsl{xrt\_dem\_iterative2.pro} routine \citep[][]{weber:04, golub:04}, modified for use with the AIA passbands. The validation 
of this method can be found in the appendix of \cite{cheng:12}. The EM was derived at 
slightly later time than the images shown in Fig.~\ref{aia_all} as some of the images were saturated at this time.The emission 
measure maps were created by calculating a DEM in every pixel using the 6 AIA EUV filters, and then integrating the DEM over 
the indicated temperature bins in order to get the emission measure in that temperature range. Errors in the DEM calculations are estimated by calculating Monte Carlo runs on the data using intensity values varied normally 
by the sigma error. This sigma error is estimated by using an empirical formula that approximates Poisson statistics for low count rates, but 
approaches Gaussian statistics for high count rates \citep[see e.g.,][]{geh_1986}. Examining the Monte Carlo runs shows that the errors in 
the DEMs are higher where the signal in several of the passbands is very low, and where the temperatures are outside the range covered 
well by the AIA passbands (i.e. $<$1 MK and $>$ 20 MK).

The EM maps shown in Fig.~\ref{em} suggest that the erupting prominence is highly multi-thermal and fairly dense at all temperatures. It should 
also be noted that the EM is high at the higher temperature bands (e.g. 10-13 MK; the higher red colours at the log EM scale), while comparatively low 
at typical coronal temperatures (e.g., 1-2 MK; the moderate yellow colour at the log EM scale) at the central kink location. This supports our premises of 
reconnection generated heating which appears in form of DEM in various temperature bands.

The EM maps also reveal a hot overlying arched structure, which might be representative of an erupting flux rope. Such hot 
erupting structures have been seen previously by \cite{revees:11,cheng:11,cheng:12,cheng:13} where they were interpreted as a 
heated erupting flux rope. However, we do not see any signature of the twist or writhe in this hot overlying flux-rope as previously 
observed by \cite{liu_2010} and \cite{zhang_2012}. Therefore, it may be plausible that this overlying structure is due to the heating 
at the current sheet between the kinked flux rope and the overlying field (T{\"o}r{\"o}k et al. 2004). Observations of such a rare heating 
scenario is possible, most likely  due to the narrow EUV passbands of SDO/AIA and availability of the hot channels.

\begin{figure}
\centering
\includegraphics[width=1.0\textwidth]{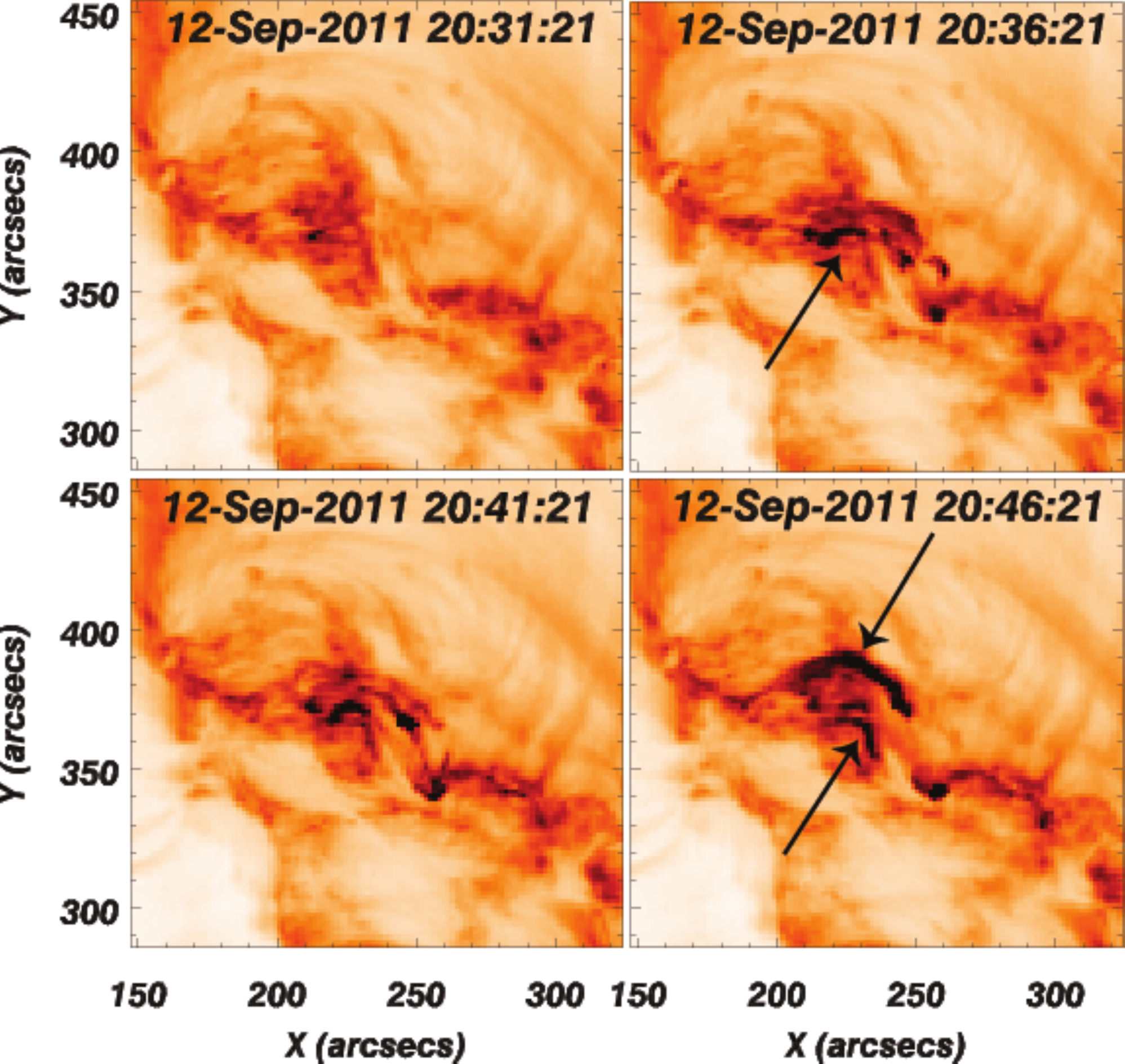}
\caption{Sequence of SECCHI 195 images showing on disk evolution of the erupting filament. \label{secchi}}
\end{figure}

As shown in Figures~\ref{aia304} \& \ref{aia_all}, the observation reported here is above the limb from the vantage point of SDO. In order to get 
an on-disk perspective, we have used the observations recorded by STEREO-b. Figure~\ref{secchi} displays images recorded by 
SECCHI using its 195~{\AA} passband. The spatial and temporal resolution of SECCHI 
images are not as good as that of AIA. However, the images shown in Figure~\ref{secchi} capture the essential dynamics of the erupting 
prominence. The arrow in the top right image show the filament in question, which is rising up. The two arrows in the bottom right panel shows 
that the filament has broken into two parts (shown by two arrows). The bottom part of the filament shows cusp shaped structure. The top 
of the cusp would correspond to the location of reconnection and enhanced heating, which is captured in the multi-wavelength observations 
by AIA filters (see Fig.~\ref{aia_all}). Observations of cusp-shaped structure is one of the most important signatures of magnetic 
reconnections \citep[see e.g.][]{fa_1996}.

\section{Summary and Discussion} \label{conc}
We have studied complete evolution of a prominence [from early phase evolution to eruption] which shows partial eruption using 
the excellent observations recorded by AIA on board SDO. The prominence show SR phase during the early phase, followed 
by a FR. The start of the SR phase coincided with two small brightenings located near the foot points. The prominence 
shows kinking motion almost at the time of transition from SR to FR. The enhanced brightening was seen at the location of 
kink, which is seen in all the wavelengths, suggestive of the presence of multi-thermal structures, a premise borne out by the emission 
measure analysis. The material above the kink location has been observed to be ejected outward (see Fig 3 and associated animation) and the plasma 
below drains down towards the surface.

To the best of our knowledge, this is the first observation where all the predicted characteristics of partial ejection, including transition 
from SR-to-FR, kinking, formation of central vertical current sheet, and splitting of filament preceded by magnetic reconnection 
has been detected in a single event, thanks to the unique time and spatial resolution of AIA. The multi-temperature coverage of the AIA 
allows us to study the temperature structure of the erupting prominences and study the heating and cooling of plasma during erupting 
prominences. 

These observations provide strong and comprehensive evidence in favour of the model of partially erupting flux ropes. 
\cite{gibson_2008} found that the magnetic connectivity, twist, orientation, and topology of the ejected portion of a 
partially-erupting flux-rope differs significantly from its pre-eruption state. While this idealized simulation cannot capture the 
details of the observations presented in this paper, the fact that the observations follow the predicted characteristics of that 
simulation argue that a significant part of the magnetic twist remains behind in the lower portion of a rope that does not 
erupt; therefore, this region is likely to experience further eruptions. It should also be noted that the observed kink is a 
configuration of the magnetic flux rope that enables the reconnection but does not necessarily manifest the evolution of 
the kink instability \citep[][]{if_2007, Tetal_2010, liu_2012}.

\acknowledgments{}
We acknowledge the suggestions of the referee that improved our manuscript.
We thank Yuhong Fan for reviewing the article and providing comments. DT acknowledges the support from DST under Fast Track Scheme 
(SERB/F/3369/2012- 2013). K.K.R. acknowledges support from contract SP02H1701R from Lockheed-Martin to SAO and NSF SHINE grant 
AGS-1156076 to SAO. The National Center for Atmospheric Research is funded by the National Science FoundationThe AIA data are courtesy of 
SDO (NASA) and the AIA consortium. The STEREO SECCHI data are produced by an international consortium of the Naval Research Laboratory 
(USA), Lockheed Martin Solar and Astrophysics Lab (USA), NASA Goddard Space Flight Center (USA), Rutherford Appleton Laboratory (UK), 
University of Birmingham (UK), Max-Planck-Institut fŸr Sonnensystemforschung (Germany), Centre Spatiale de Lige (Belgium), Institut d'Optique 
ThŽorique et AppliquŽe (France), and Institut d'Astrophysique Spatiale (France).


\end{document}